\pdfoutput=1 

\documentclass{article}
\usepackage[T1]{fontenc} 
\usepackage[utf8]{inputenc} 
\usepackage{ismir,amsmath,cite,url}
\usepackage{graphicx}
\usepackage{color}
\usepackage{multirow}
\usepackage{float}
\usepackage{kotex}
\usepackage{makecell}
\usepackage[bookmarks=false,hidelinks]{hyperref}

\title{A BI-DIRECTIONAL TRANSFORMER FOR MUSICAL CHORD RECOGNITION }

\multauthor
{Jonggwon Park \hspace{1cm} Kyoyun Choi \hspace{1cm} Sungwook Jeon} { \bfseries{Dokyun Kim \hspace{1cm} Jonghun Park}\\
Department of Industrial Engineering \& Center for Superintelligence,\\
Seoul National University, Seoul, Republic of Korea\\
{\tt\small \{jayg996, andyhome1907, wookee3, kdk01, jonghun\}@snu.ac.kr}
}
\def\authorname{Jonggwon Park, Kyoyun Choi, Sungwook Jeon, Dokyun Kim and Jonghun Park}

\begin{document}

\maketitle

\begin{abstract}
Chord recognition is an important task since chords are highly abstract and descriptive features of music.
For effective chord recognition, it is essential to utilize relevant context in audio sequence. 
While various machine learning models such as convolutional neural networks (CNNs) and recurrent neural networks (RNNs) have been employed for the task, most of them have limitations in capturing long-term dependency or require training of an additional model.

In this work, we utilize a self-attention mechanism for chord recognition to focus on certain regions of chords.
Training of the proposed bi-directional Transformer for chord recognition (BTC) consists of a single phase while showing competitive performance.
Through an attention map analysis, we have visualized how attention was performed.
It turns out that the model was able to divide segments of chords by utilizing adaptive receptive field of the attention mechanism.
Furthermore, it was observed that the model was able to effectively capture long-term dependencies, making use of essential information regardless of distance.

\end{abstract}

\section{Introduction}\label{sec:introduction}

The goal of chord recognition task is to output a sequence of time-synchronized chord labels when a raw audio recording of music is given as input.
Chords are highly abstract and descriptive features of music that can be used for a variety of musical purposes, including automatic lead-sheet creation for musicians, cover song identification, key classification and music structure analysis\cite{Lee:2006, Bello:2007, Pauwels:2013}.
Since manual chord annotation is labor intensive, time consuming and requires expert knowledge, automatic chord recognition system has been an active research area within the music information retrieval community.

Automatic chord recognition is challenging due to the fact that 1) not all the notes played are necessarily related to the chord of the moment and 2) simple one-hot encoding of chord labels cannot represent the inherent relationship between different chords. 
Most traditional automatic chord recognition systems consist of three parts: feature extraction, pattern matching and chord sequence decoding.
The most common strategy was to rely on hidden Markov models (HMMs) \cite{Baum:1966} for sequence decoding.
Recently, many studies have explored various deep neural networks such as convolutional neural networks (CNNs) or recurrent neural networks (RNNs)\cite{LeCun:2015} for chord recognition.

Recently, a novel attention-based network architecture named Transformer was proposed in \cite{Vaswani:2017}.
It performs well without any recurrence or convolution and the use of Transformer has become popular in various domains.
For example, a bi-directional Transformer model called BERT achieved state-of-the-art results on eleven natural language processing (NLP) tasks\cite{Devlin:2018}.
In the domain of music, \cite{Huang:2018} applied Transformer to a music generation task and succeeded in creating music with complex and repetitive structure.

In this paper we propose BTC (Bi-directional Transformer for Chord recognition).
In contrast to the other chord recognition models that depend on training of separate feature extractors or adopting additional decoders such as HMMs or Conditional Random Fields (CRFs) \cite{Lafferty:2001}, BTC requires only a single training phase while being able to obtain results comparable to them.
We also visualize how the model works through attention maps.
The attention maps demonstrate that BTC is able to 1) divide segments of chords by utilizing its adaptive receptive field and 2) capture long-term dependencies.


\section{Related Work}\label{sec:related_work}

\subsection{Automatic Chord Recognition}\label{subsec:automatic_chord_recognition}
In the past, most automatic chord recognition systems were divided into three parts: feature extraction, pattern matching and chord sequence decoding. 
After applying transformation such as short-time Fourier transform or constant-q transform (CQT) to an input audio signal, features are extracted from the resulting time-frequency domain.
Some examples of such hand-crafted features include chroma vectors and the "Tonnetz"\cite{Euler:1739} representation.
For pattern matching and chord sequence decoding, Gaussian mixture models with feature smoothing\cite{Cho:2014a,Cho:2011} and HMMs\cite{Sheh:2003, Ueda:2010} have been the most popular choices, respectively. 

With the recent wide acceptance of deep learning in research communities, there have been many studies applying it to chord recognition task in various ways.
The very first deep-learning-based chord recognition system was proposed by \cite{Humphrey:2012a} where they trained a CNN for major-minor chord classification.
Attempts to apply deep learning to feature extraction include \cite{Humphrey:2012b} and \cite{Korzeniowski:2016a}, where the former employed a CNN to extract Tonnetz features from audio data and the latter adopted a deep neural network (DNN) to compute chroma features.
CNN and HMM were combined for chord recognition in \cite{Humphrey:2015} and \cite{Zhou:2015}.

In addition to CNN, another popular network architecture for chord recognition is RNN.
\cite{Lewandowski:2013} and \cite{Sigtia:2015} explored an RNN as chord sequence decoding method, relying on deep belief network and a DNN, respectively.
Another branch of RNN-based chord recognition systems utilize a language model which predicts only the sequence of chords without considering their durations. 
This might be helpful when the number of chord labels is large (e.g. large vocabulary type, explained in Section \ref{subsec:data_preproces}). 
A large-scale study of language models for chord prediction was conducted in \cite{Korzeniowski:2018a}.
Without audio data, the authors trained just a language model with the chord progression data only and showed that RNNs outperformed N-gram models.
In their succeeding work\cite{Korzeniowski:2018b}, they combined the RNN-based harmonic language model with a chord duration model to complete the chord recognition task.

Another RNN-based approach is presented in \cite{Wu:2019} which trained a CNN feature extractor with large MIDI (Musical Instrument Digital Interface) data and combined BLSTM (Bi-directional Long-Short Term Memory) with CRF for sequence decoder.
This BLSTM-CRF model achieved good performance but has a drawback that its training procedure involves complex MIDI pre-training.
The model that we propose, on the other hand, is much simpler to train.

\subsection{Attention-based Models}\label{subsec:attention models}
The attention mechanism, first introduced by \cite{Bahdanau:2014}, can be described as computing an output vector when query, key and value vectors are given.
In sequence modelling tasks such as machine translation, query and key correspond to certain elements of the target sequence and the source sequence respectively. 
Each key has its own value.
The output is computed as a weighted sum of the values where the weights are computed from the query and key.
Self-attention refers to the case when query, key and value are computed from the same input.

Transformer is an attention-based network that relies on attention mechanism only and does not include recurrent or convolutional architecture. 
Utilizing multi-head attention together with position-wise fully-connected feed-forward network, it showed significantly faster training speed and achieved better performance than recurrent or convolutional networks for translation tasks.

Transformer used scaled dot-product as an attention function:
\begin{equation}\label{attention}
Attention(Q,K,V) = softmax( \frac{QK^{T}}{\sqrt{d_{K}}} ) V
\end{equation}
where \(Q\), \(K\) and \(V\) are matrices of query, key and value vectors respectively, and \(d_K\) is the dimension of key.

The use of Transformer has become very popular, achieving the state-of-the-art results in various domains.
A well-known example is bi-directional encoder representations from Transformers (BERT)\cite{Devlin:2018}.
BERT is a pre-training model based on masked language model for language representations that achieved state-of-the-art results on eleven NLP tasks.
In the domain of music, \cite{Huang:2018} proposed music Transformer for symbolic music generation.
Music Transformer employed relative attention to capture long-term structure effectively, which resulted in music compositions that are both qualitatively and quantitatively better structured than existing music generation models.

\begin{figure*}[t]
 \centerline{
 \includegraphics[width=2.0\columnwidth]{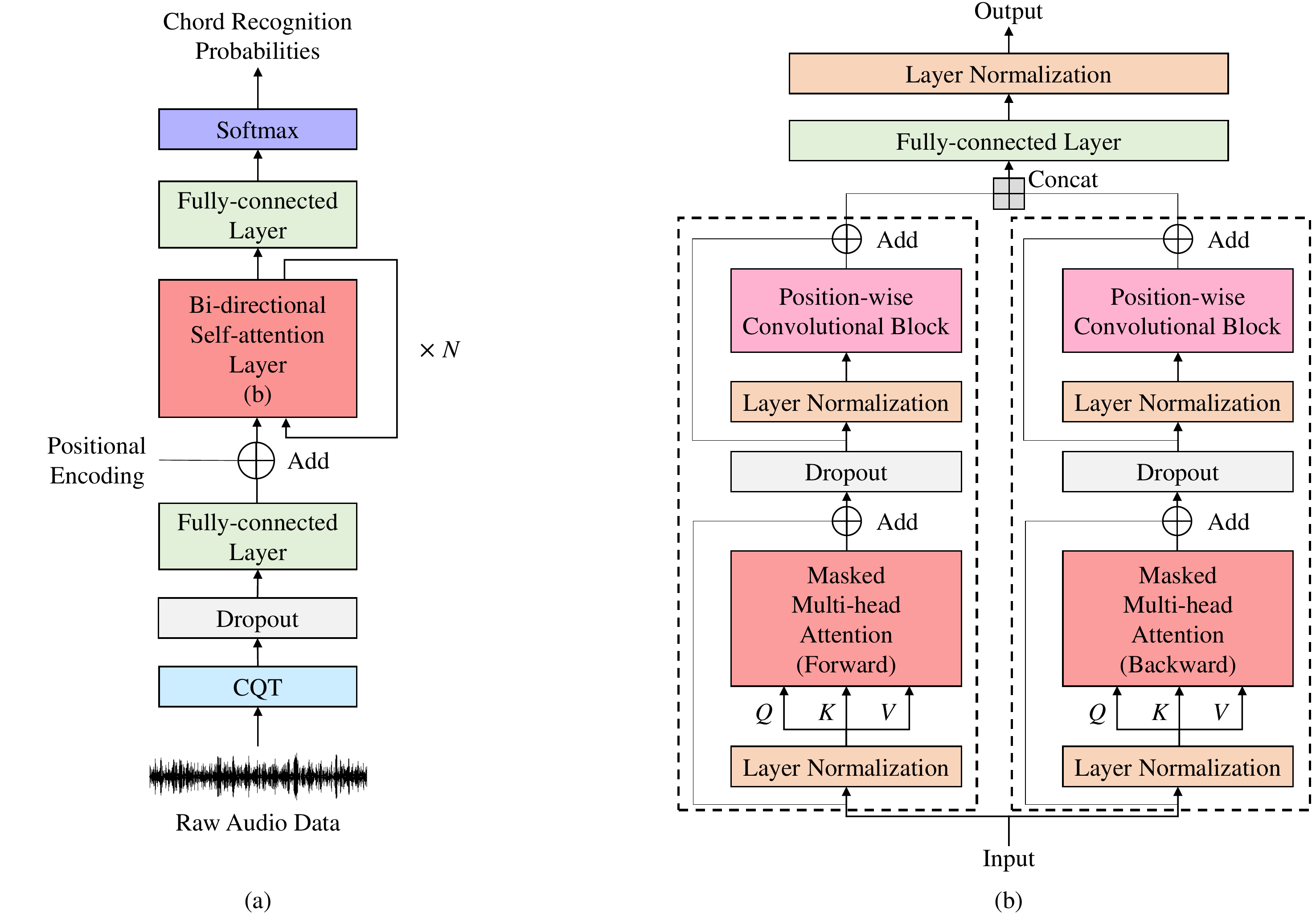}}
 \caption{Structure of BTC. (a) shows the overall network architecture and (b) describes the bi-directional self-attention layer in detail. Dotted boxes indicate self-attention blocks.}
 \label{fig:model}
\end{figure*}

\section{Bi-directional Transformer for Chord recognition}\label{sec:model}

\subsection{Bi-directional Transformer}\label{subsec:model1}

Making use of appropriate surrounding frames is essential for successful chord recognition\cite{Cho:2014b,Cho:2011}.
This context-dependent characteristic of the task is the motivation for applying the self-attention mechanism.
With some modification to the original Transformer architecture, we present a bi-directional Transformer for chord recognition (BTC).\footnote{\url{https://github.com/jayg996/BTC-ISMIR19}}

The structure of BTC is shown in \figref{fig:model}. 
The model consists of bi-directional multi-head self-attentions, position-wise convolutional blocks, a positional encoding, layer normalization \cite{layernorm}, dropout \cite{Srivastava:2014} and fully-connected layers. 
The model takes a CQT feature of 10 second audio signal (\secref{subsec:data_preproces}) as input.
The results of adding positional encoding are given as input to two self-attention blocks with different masking directions, indicated as dotted boxes in \figref{fig:model}(b).
The outputs are concatenated and are fed into a fully-connected layer so that the output size is the same as the original input. 
A stack of \( N \) bi-directional self-attention layers is followed by another fully-connected layer that outputs logit values.
The size of the logit values is the same as the number of chord labels. 
These logits are used to predict the chord and calculate the loss. 

The loss function is a negative log-likelihood and all the model parameters are trained to minimize the loss given by the following equation \eqref{loss}. 

\begin{equation}\label{loss}
L = - \sum_{t=1}^{T} \sum_{c \in{V}} y_c(t) log( \hat{y_{c}} (t) )
\end{equation}
\( T \) is the number of total time frames and \(V \) is the chord label set. 
\( y_c(t) \) is 1 if the reference label at time \( t \) is \( c \) and 0 otherwise.
\( \hat{y_{c}} (t) \) is the output of the model, representing the probability of the chord at time \( t \) being \( c \).

\subsubsection{Bi-directional Multi-head Self-attention}\label{subsubsec:attention_module}

BTC employs multi-head self-attention as in the original Transformer.
For each time frame, the input features are split into \( n_h \) pieces and provided as input to the multi-head self-attention with the number of heads, \( n_h \).
Given \(I\) as an input matrix, the multi-head self-attention can be computed as \eqref{multi-head}:

\begin{equation}\label{multi-head}
Multihead = Concat(head_1, ..., head_{n_h})W_{O}
\end{equation}
\( Q_j = (IW_{Q})_j, K_j = (IW_{K})_j\) and \( V_j = (IW_{V})_j \) are given as input to the attention function \eqref{attention} to produce \( head_j \) for \( j = 1, ..., n_h\). 
\(W_Q, W_K\) and \(W_V\) are fully-connected layers that project the input to the dimension of \(Q,K\) and \(V\), respectively.
\(W_O\) is also a fully-connected layer that projects the concatenated output of dimension (\(n_h \times d_{V_j}\)) to the dimension of the final output. 
Dropout is applied to the softmax output weights when computing each \(head_j\).

In BTC, self-attention can be interpreted as determining how much attention to apply to the value of the key time frame when inferring the chord of the query time frame.
To prevent the loss of information due to the attention being performed to the entire input at once, we employed bi-directional masking.
The forward / backward direction refers to masking all the preceding / succeeding time frames.
The same masked multi-head attention module as the Transformer decoder was adopted. 
The bi-directional structure enables BTC to fully utilize the context before and after the target time frame.

Since the multi-head attention is performed on every time frame in the sequence, information about the order of the sequence is lost.
We employed the same solution proposed by Transformer to address this issue: adding positional encoding results to the input, which are obtained by applying sinusoidal functions to each position.
Since relative positions between two frames can be expressed as a linear function of the encodings, positional encoding helps the model learn to apply attention via relative positions.

\subsubsection{Position-wise Convolutional Block} \label{subsubsec:convolution}
To utilize the adjacent feature information in a time frame, we replaced the position-wise fully-connected feed-forward network from the original Transformer architecture with a position-wise convolutional block.
The position-wise convolutional block consists of a 1D convolution layer, a ReLU (Rectified Linear Unit) activation function and a dropout layer, where the whole sequence of layers is repeated \( n_C \) times. 
Input and output channel size were identical to keep the feature size and sequence length constant. 
With the position-wise convolutional block, we anticipate to search the boundary and smooth the chord sequence by exploring adjacent information at each time frame. 

\subsection{Self-attention in Chord Recognition}\label{subsec:whyapply}
For chord recognition, it is important to utilize not only the information from the target time frame but also from other related frames, which we call the context.
The network architectures such as CNNs or RNNs can also explore the context, but self-attention is more suitable for the task because of the following reasons.

\begin{figure}[ht]
 \centerline{
 \includegraphics[width=1.0\columnwidth]{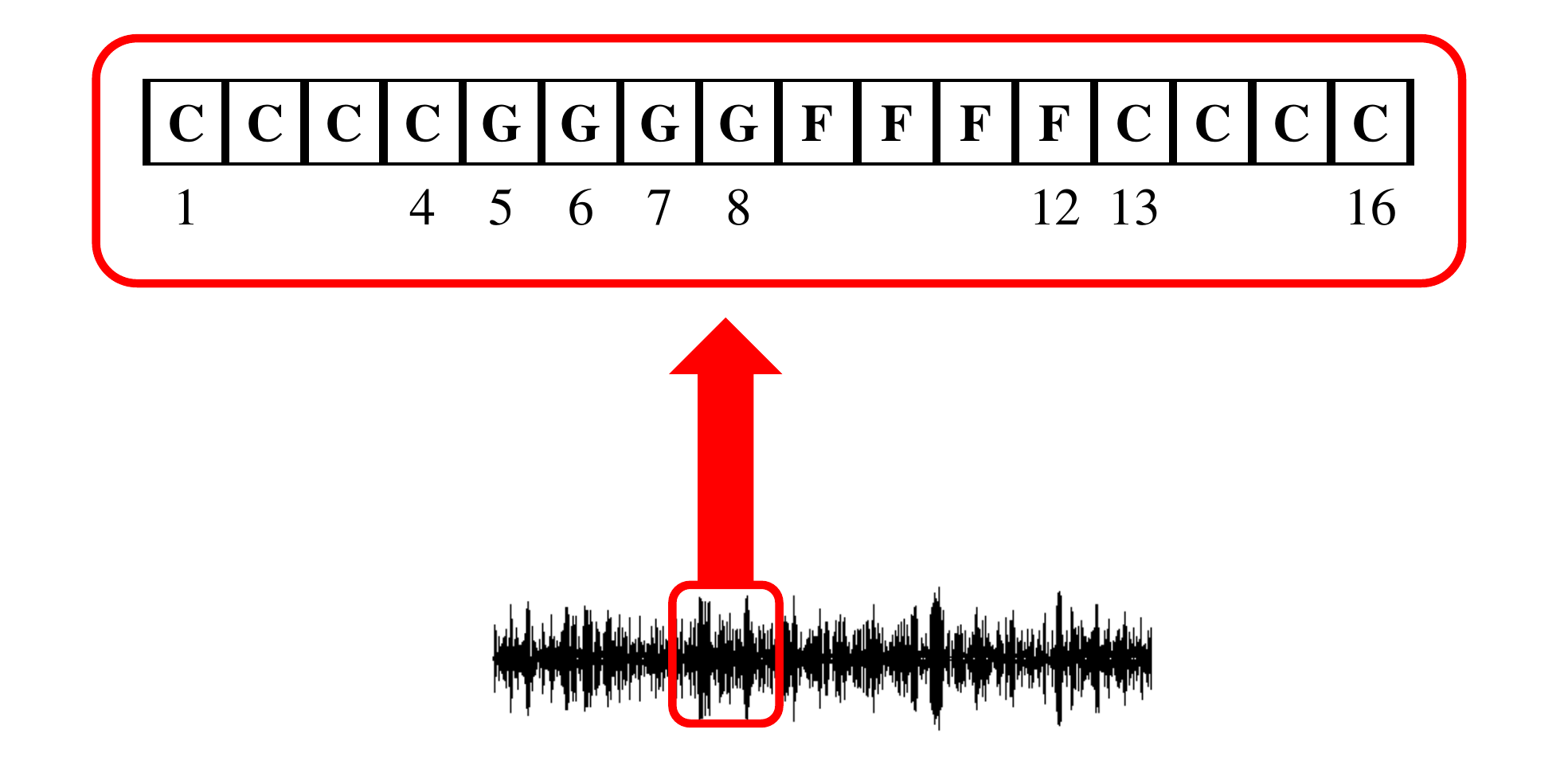}}
 \caption{Chord sequence example}
 \label{fig:example}
\end{figure}

First, self-attention has selective usage of attention.
In other words, the receptive field can be adaptive unlike CNNs where the kernel size is fixed.
For example, assume that the labels for 16 frames are Cs for the first four frames, Gs and Fs for the next eight frames and Cs for the last four frames (see \figref{fig:example}).
Consider the situation of recognizing Gs in frames 5 to 8.
As for a CNN with kernel size of 3, when recognizing the chord of frame 7, the receptive field (frame 6 to 8) would be informative enough since all the frames contain the same chord.
However, when inferring frame 5, the receptive field of frame 4 to 6 contains not only G but also C.
With self-attention, on the other hand, the model can pay attention to the section of frame 5 to 8 regardless of the target frame's position.

Another advantage of attention mechanism is its ability to capture long-term dependency effectively.
RNNs can also utilize distant information but direct access is not possible.
For CNNs, there are two ways to access distant frames: by stacking layers in depth or by increasing the kernel size.
The former has the same drawback as RNNs and the latter has the disadvantage that the weight sharing becomes less effective.
Unlike these, self-attention has direct access to other frames no matter how far they are.
Specifically, when recognizing the chord of frame 13, performing attention to first four frames would be helpful since they all contain C.
With RNNs or deep CNNs, information that the first four frames were C would inevitably be diluted while passing through frames 5 to 12.

\begin{table*}[t]
 \begin{center}
\begin{tabular}{c|cc|ccccccc}
\Xhline{3\arrayrulewidth}
\multirow{2}{*}{Model} & \multicolumn{2}{c|}{maj-min label type} & \multicolumn{7}{c}{large vocabulary label type}\\ \cline{2-10}
& Root & Maj-min &Root  & Thirds & Triads & Sevenths & Tetrads & Maj-min & MIREX \\ \Xhline{3\arrayrulewidth}
CNN & \normalsize{83.6}\tiny{\(\pm 1.3\)} & \normalsize{81.8}\tiny{\(\pm 1.2\)} & \normalsize{83.5}\tiny{\(\pm 1.4\)} & \normalsize{80.4}\tiny{\(\pm 1.2\)} & \normalsize{75.5}\tiny{\(\pm 0.6\)} & \normalsize{71.5}\tiny{\(\pm 1.9\)} & \normalsize{65.2}\tiny{\(\pm 1.0\)} & \normalsize{81.9}\tiny{\(\pm 1.4\)} & \normalsize{79.8}\tiny{\(\pm 0.7\)}  \\
CNN+CRF \cite{Korzeniowski:2016b} & \textbf{84.0}\tiny{\(\pm 1.3\)} & \textbf{83.1}\tiny{\(\pm 1.4\)} & \textbf{83.7}\tiny{\(\pm 1.5\)} & \textbf{81.1}\tiny{\(\pm 1.4\)} & \textbf{76.3}\tiny{\(\pm 0.8\)} & 71.3\tiny{\(\pm 1.9\)} & \textbf{65.7}\tiny{\(\pm 1.6\)} & 82.1\tiny{\(\pm 1.5\)} & \textbf{81.8}\tiny{\(\pm 1.1\)}\\ 
CRNN \cite{McFee:2017} & 83.4\tiny{\(\pm 0.8\)} & 82.3\tiny{\(\pm 0.9\)} & 82.9\tiny{\(\pm 1.1\)} & 80.1\tiny{\(\pm 1.0\)}  & 75.3\tiny{\(\pm 0.7\)} & 71.3\tiny{\(\pm 1.9\)} & 65.2\tiny{\(\pm 0.9\)} & 81.5\tiny{\(\pm 1.3\)} & 79.9\tiny{\(\pm 0.8\)} \\ 
CRNN+CRF & 83.3\tiny{\(\pm 0.8\)} & 82.3\tiny{\(\pm 1.0\)} & 82.7\tiny{\(\pm 1.2\)} & 79.7\tiny{\(\pm 0.9\)}  & 74.8\tiny{\(\pm 0.5\)} & 69.5\tiny{\(\pm 2.0\)} & 63.9\tiny{\(\pm 1.0\)} & 80.7\tiny{\(\pm 1.4\)} & 80.2\tiny{\(\pm 1.0\)} \\ \hline
BTC & 83.8\tiny{\(\pm 1.0\)} & 82.7\tiny{\(\pm 1.0\)} & 83.5\tiny{\(\pm 1.2\)} & 80.8\tiny{\(\pm 1.0\)} & 75.9\tiny{\(\pm 0.5\)} & \textbf{71.8}\tiny{\(\pm 1.7\)} & 65.5\tiny{\(\pm 0.9\)} & \textbf{82.3}\tiny{\(\pm 1.2\)} & 80.8\tiny{\(\pm 0.9\)} \\
BTC+CRF & 83.9\tiny{\(\pm 1.0\)} & \textbf{83.1}\tiny{\(\pm 1.1\)} & 83.5\tiny{\(\pm 1.2\)} & 80.7\tiny{\(\pm 1.1\)} & 75.7\tiny{\(\pm 0.5\)} & 70.7\tiny{\(\pm 2.0\)} & 64.8\tiny{\(\pm 1.1\)} & 81.7\tiny{\(\pm 1.4\)} & 81.4\tiny{\(\pm 0.9\)} \\
\hline
\end{tabular}
\end{center}
 \caption{WCSR scores averaged over the same 5 folds. Numbers next to the scores denote the standard deviations.}
 \label{tab:results} 
\end{table*}

\section{Experiments}\label{sec:experiment}

\begin{figure*}[ht]
 \centerline{
 \includegraphics[width=2.0\columnwidth]{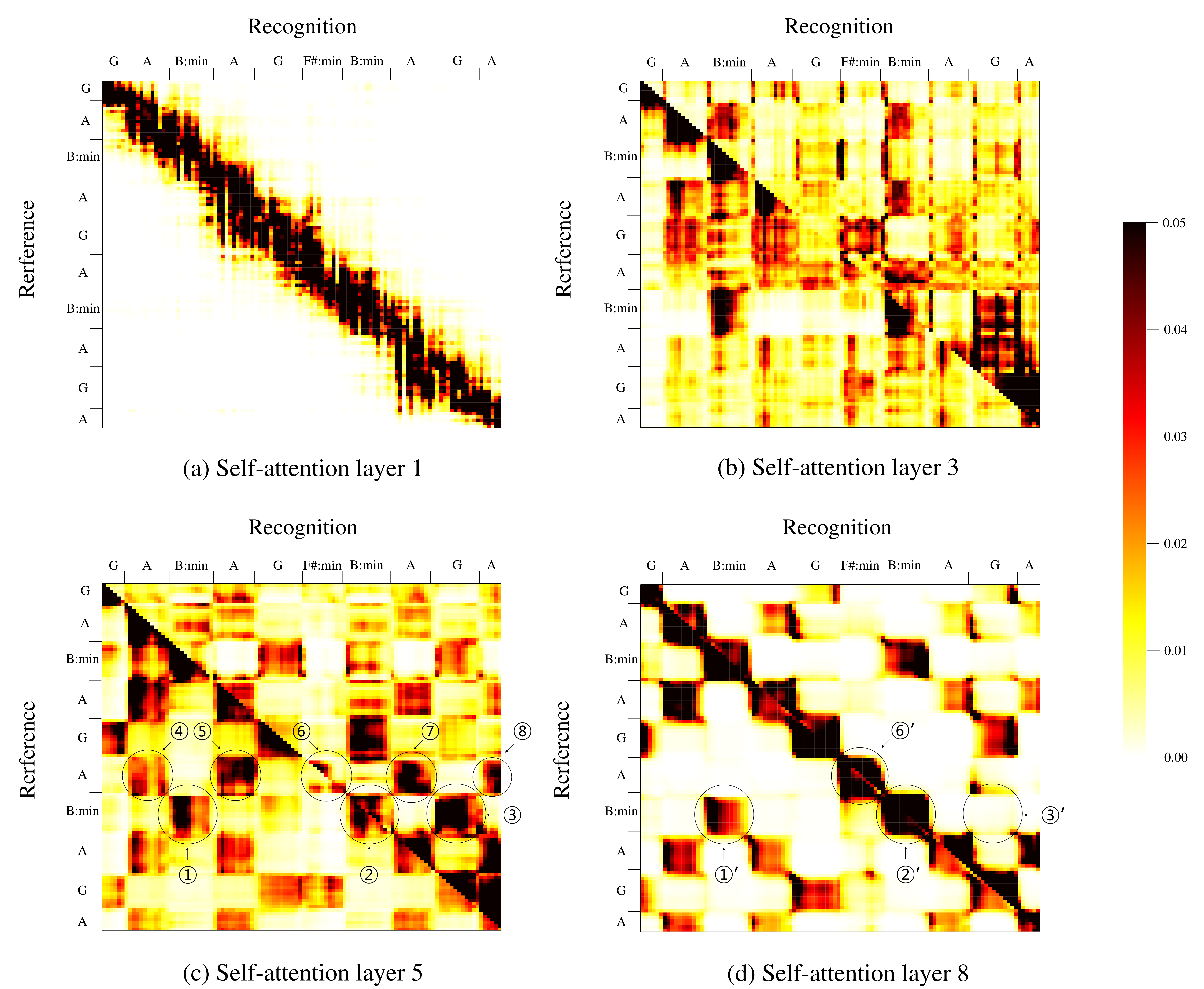}}
\caption{The figures represent the probability values of the attention of self-attention layers 1, 3, 5 and 8 respectively.
The layers that best represent the different characteristics were chosen.
The input audio is the song "Just A Girl" (0m30s \(\sim\) 0m40s) by No Doubt from UsPop2002, which was in evaluation data.}
 \label{fig:attention}
\end{figure*}

\subsection{Data and Preprocessing}\label{subsec:data_preproces}

BTC and other baseline models were evaluated on the following datasets.
A subset of 221 songs from  Isophonics\footnote{\url{http://isophonics.net/datasets}}: 171 songs by the Beatles, 12 songs by Carole King, 20 songs by Queen and 18 songs by Zweieck; Robbie Williams \cite{dataset}: 65 songs by Robbie Williams; and  a subset of 185 songs from UsPop2002\footnote{\url{https://github.com/tmc323/Chord-Annotations}}.
These datasets consist of label files that specify the start time, end time and type of the chord.
Due to copyright issue, these datasets do not include audio files.
The audio files used in this work were collected from online music service providers (e.g. Melon\footnote{\url{http://www.melon.com}}), which do not always provide the same audio files corresponding to the songs in the datasets.
Since it was not possible to get exactly the same audio files, there were subtle differences in the chord start time of the label file and audio file.
Accordingly we manually matched the labels to the audio file by shifting the whole label file back and forth, which resulted in no more than adding or deleting some “No chord” labels. 

Each 10-second-long audio signal (consecutive signals overlapping 5 seconds) was processed at the sampling rate of 22,050Hz using CQT with 6 octaves starting from C1, 24 bins per octave, and the hop size of 2048\cite{Wu:2019}.
The CQT features were transformed to log amplitude with \( S_{log} = ln (S + \epsilon) \) where \(S\) represents the CQT feature and \(\epsilon\) is an extremely small number.
After that, global z-normalization was applied with mean, variance from the training data. 

Pitch augmentation was also employed to the audio file with pyrubberband\footnote{\url{https://github.com/bmcfee/pyrubberband}} package and labels were changed with pitch variation. Pitch augmentation between -5 \(\sim\) +6 semitones were applied to all the training data.

Two different label types were used: maj-min and large vocabulary. 
The maj-min label type consists of 25 chords (12 semitones \(\times\) \{maj, min\} and “No chord”)\cite{Korzeniowski:2016b}. 
The large vocabulary label type consists of 170 chords (12 semitones \(\times\) \{maj, min, dim, aug, min6, maj6, min7, minmaj7, maj7, 7, dim7, hdim7, sus2, sus4\} and “X chord : the unknown chord”, “No chord”)\cite{McFee:2017}.
From the label files, we extracted the chord that matches the time frame of input feature and transformed it to the appropriate label type. 

\subsection{Evaluation Metric}\label{subsec:performance}

The evaluation metric was weighted chord symbol recall (WCSR) score and 5-fold cross validation was applied to the entire data.
When separating the evaluation data from the training data, there was no song included in both.
The WCSR score can be computed as \eqref{wcsr}, where \(t_c\) is the duration of correctly classified chord segments and \(t_a\) is the duration of the entire chord segments.

\begin{equation}\label{wcsr}
WCSR = \frac{t_c}{t_a} \times 100 (\%)
\end{equation}

Scores were computed with mir\_eval\cite{mireval}. Root and Maj-min scores were used for the maj-min label type.
Root, Thirds, Triads, Sevenths, Tetrads, Maj-min and MIREX scores were used for the large vocabulary label type.
To calculate the score with mir\_eval, the chord recognition results were converted into label files.

\subsection{Results}\label{subsec:results}
Specific hyperparameters of BTC are summarized in \tabref{tab:parameters}.
The hyperparameters with the best validation performance were obtained empirically after applying in 5-fold cross validation.
Adam optimizer\cite{DBLP:journals/corr/KingmaB14} was used with initial learning rate of \( 10^{-4}\).
Learning rate was decayed with rate 0.95 when validation accuracy did not increase.
Training was stopped if the validation accuracy did not improve for over 10 epochs.

Since existing studies of chord recognition were evaluated on different datasets, it is difficult to say that a particular model is the state-of-the-art.
Among the models that were trainable with our datasets, we chose three baseline models with good performance: CNN, CNN+CRF and CRNN.
CNN is a VGG\cite{Simonyan:2014}-style CNN and CNN+CRF has an additional CRF decoder\cite{Korzeniowski:2016b}.
CRNN is a combination of CNN and gated recurrent unit\cite{DBLP:journals/corr/ChungGCB14}, named "CR2" in \cite{McFee:2017}. 
The input was preprocessed as mentioned in \secref{subsec:data_preproces} for BTC and CRNN.
For CNN+CRF and CNN, a single label was estimated with a patch of 15 time frames, in a similar way to \cite{Korzeniowski:2016b}.

\begin{table}[b]
\centering
\begin{tabular}{|c|l|c|}
\hline
\multirow{3}*[-1.5ex]{
\begin{tabular}[c]{@{}c@{}}Bi-directional\\ self-attention\\ layer\end{tabular}} 
& layer repetition (\( N \)) & \small{\{1, 2, 4, \textbf{8}, 12\}}   \\ \cline{2-3} 
& self-attention heads (\( n_h \))& \small{\{1, 2, \textbf{4}\}}   \\ \cline{2-3} 
& \begin{tabular}[c]{@{}l@{}}dimension of \(Q\), \(K\), \(V\) \\ and all the hidden layers\end{tabular} & \small{\{64, \textbf{128}, 256\}} \\ \hline
\multirow{4}{*}{\begin{tabular}[c]{@{}c@{}}Position-wise\\ convolutional\\ block\end{tabular}}   
& block repetition (\( n_C \)) & \small{\textbf{2}}   \\ \cline{2-3} 
& kernel size & \small{\textbf{3}} \\ \cline{2-3} 
& stride & \small{\textbf{1}}   \\ \cline{2-3} 
& padding size & \small{\textbf{1}}   \\ \hline
Dropout              
& dropout probability & \small{\{\textbf{0.2}, 0.3, 0.5\}} \\ \hline \end{tabular}
\caption{Hyperparameters of BTC. Hyperparameters with the best validation performance are shown in bold. }
\label{tab:parameters}
\end{table}

\tabref{tab:results} shows the performance comparison results of the baseline models and BTC for two label types. 
The best value for each metric is represented in bold.
Among the models without a CRF decoder, BTC showed the best performance for all metrics.
Including models with a CRF decoder, CNN+CRF obtained the best result in most of the metrics.
Still, BTC shows comparable performance to CNN+CRF, performing better in Sevenths and Maj-min metrics for the large vocabulary label type. 

The main purpose of training a CRF decoder is to smooth the predicted chord sequences that are often fragmented.
The performances of CRNN+CRF and BTC+CRF are also presented in \tabref{tab:results} for comparison.
Performance improvements due to the introduction of CRFs are evident in CNN but not in BTC and CRNN.
This indicates that outputs of CNN were fragmented and an additional decoder training is necessary for better performance.
On the other hand, BTC and CRNN can be trained with only CQT features and chord labels.
That is, BTC requires only a single training phase while achieving the performance comparable to that of CNN+CRF. 

\subsection{Attention Map Analysis }\label{subsec:attention_analysis}

Attention maps demonstrate that each self-attention layer has different characteristics. 
\figref{fig:attention} shows the attention map of self-attention layers 1, 3, 5 and 8, trained with the maj-min label type.
The lower / upper triangle of each attention map represents the attention probability of the forward / backward direction self-attention layer.
The labels of the vertical axis and the horizontal axis are the reference chord and the chord recognition result of the target time frame, respectively.
The cell of \(i\)-th row and \(j\)-th column represents the attention probability to the \(j\)-th time frame when inferring the chord of the \(i\)-th time frame.

At the first self-attention layer, only neighboring frames are used to construct the representation of the target frame. 
For the third layer, the attention is widely spread over all time frames, yet still with higher probabilities for nearby frames than distant frames. 
At the fifth layer, several adjacent time frames form a group, which appears in a rectangular region in the attention map. 
This means that the model divides the whole input into some sections, which is possible due to the adaptive receptive field.
The network focuses only on a few important sections to identify the target frame, regardless of the distance between section and the frame.
Unlike the fifth layer, attention is more dense in certain regions at the eighth layer.
In particular, the boundary of the high probability region matches that of the final recognition result. 

Specifically, at the fifth layer in \figref{fig:attention}(c), the reference chord for region ② is B:min.
Region ① shares the same reference chord B:min and the network assigns high attention probabilities to region ① for time frames in region ②.
This phenomenon is similar in layer 8 between ①′ and ②′(\figref{fig:attention}(d)), which results in the correct final chord recognition of B:min.
In contrast, for region ③ where the reference chord is G, the attention probability is high at layer 5 but not for region ③′ at layer 8.
This can be attributed to G and B:min sharing two notes in common, since G and B:min consist of (G,B,D) and (B,D,F\#) respectively. 
In other words, attention at layer 5 can be seen as attention to partial features of chords sharing the same notes. 
None the less, the final recognition result after the last layer is not G but B:min.
This is possible because of the multi-head attention structure: the other heads might lower the attention probability even if the attention to a wrong chord is active, leading to the correct result.

On the other hand, there are cases where the recognition results are wrong in a similar situation.
The reference chord for regions ⑥ and ⑥′ is A.
At layer 5, the attention mechanism seems to work well with high attention probabilities to region ④,⑤,⑦ and ⑧, where the reference chords are all As.
However, the attention to those regions cannot be seen at the last layer, and the final recognition result is not A but F\#:min.
This recognition failure can be regarded as a result of two notes of F\#:min (F\#,A,C\#) overlapping with A (A,C\#,E).

To summarize, for each target frame in the input audio, the model uses only neighboring frames at first.
At the middle layers, the model gradually broadens the receptive field and selectively focuses on time frames with characteristics similar to that of the target frame.
Finally, at the last layer, the attention is performed on only essential information for chord recognition.

\section{Conclusion}\label{sec:conclusion}
In this paper, we presented bi-directional Transformer for chord recognition (BTC).
To the best of our knowledge, this paper was the first attempt to apply Transformer to chord recognition.
The self-attention mechanism was appropriate for the task that attempts to capture long-term dependency by effectively exploring relevant sections.
BTC has an advantage in that its training procedure is simple and it showed results competitive to other models in most of the evaluation metrics. 
Through the attention map analysis, it turned out that each self-attention layer had different characteristics and that the attention mechanism was effective in identifying sections of chords that were crucial for chord recognition.


\section{Acknowledgements}\label{sec:acknowledgements}
This work was supported by Kakao and Kakao Brain corporations.

\bibliography{ISMIRtemplate}

\end{document}